\documentclass{article}
\usepackage{spconf,amsmath,graphicx}
\usepackage{pifont}
\usepackage{multirow}
\newtheorem{theorem}{Theorem}

\title{HUT: Enabling High-UTility, Batched Queries\\ under Differential Privacy Protection for Internet-of-Vehicles}
{
\name{Junyu Liu$\textsuperscript{1}$, Wangkai Jin$\textsuperscript{1}$, Zhenyong He$\textsuperscript{2}$, Xiaoxing Ming$\textsuperscript{1}$, Yicun Duan$\textsuperscript{1}$, Zeyu Xiong$\textsuperscript{1}$,  Xiangjun Peng$\textsuperscript{1}$}
\address{\textsuperscript{1}User-Centric Computing Group,University of Nottingham Ningbo China \\ 
        \textsuperscript{2}Department of Computer Science, Johns Hopkins University \\
         https://unnc-ucc.github.io/
        }
}

\begin{document}
\maketitle
\begin{abstract}
The emerging trends of Internet-of-Vehicles (IoV) demand centralized servers to collect/process sensitive data with limited computational resources on a single vehicle. Such centralizations of sensitive data demand practical privacy protections. One widely-applied paradigm, Differential Privacy, can provide strong guarantees over sensitive data by adding noises. However, directly applying DP for IoV incurs significant challenges for data utility and effective protection. We observe that the key issue about DP-enabled protection in IoV lies in how to synergistically combine DP with special characteristics of IoV, whose query sequences are usually formed as unbalanced batches due to frequent interactions between centralized servers and edge vehicles.

To this end, we propose HUT, a new algorithm to enable \underline{H}igh \underline{UT}ility for DP-enabled protection in IoV. Our key insight is to leverage the inherent characteristics in IoV: the unbalanced batches. Our key idea is to aggregate local batches and apply Order Constraints, so that information loss from DP protection can be mitigated. We evaluate the effectiveness of HUT against the state-of-the-art DP protection mechanisms. The results show that HUT can provide much lower information loss by 95.69\% and simultaneously enable strong mathematically-guaranteed protection over sensitive data.
\end{abstract}
%
%\begin{keywords}
%Differential Privacy, Internet of Vehicle, Data Query
%\end{keywords}
%
\section{Introduction}
\label{sec:intro}

Internet-of-Vehicles (IoV) is considered a promising solution to resolve the limited computational power of a single vehicle. IoV centralizes statistics from edge vehicles in a power server for computation. However, the centralizations of these statistics may contain sensitive data, and substantially requires privacy protections in practice~\cite{privacy1,privacy2,hci22/face2statistics,automotiveui20/face2multimodal,automotive21/facial-expressions, hci21/dbscan-driving-styles}. Differential Privacy (DP)~\cite{dp}, as the dominant mechanism for privacy-preserving computation, provides strong guarantees over sensitive data by statistically adding noises. Though DP is a promising solution for the privacy concerns in IoV, directly applying DP for IoV incurs significant challenges in terms of both data utility and effective protection. 

%In the era of Internet of Vehicles (IoV),  intelligent vehicles are viewed as edge devices that support various services (e.g. driver status monitoring, lane changing management). Most services require data-driven decision-makings, which demands frequent data transmissions among centralized servers to collect/retrieve/process user data, via "perceptors" equipped in the vehicles. Such behavior can increase the risk of data leakage, as adversaries  have more opportunities to attack and extract private data from the database. To this end, various privacy protection methods have been proposed to protect user privacy in IoV, such as \cite{privacy1,privacy2}. Among them, Differential Privacy (DP) \cite{dp}, an advanced and effective privacy-preserving technique that enables users to get expected query responses on randomized dataset (i.e. add Laplace Noise for protection), has been widely adopted to tackle such issues.

We address the key issue behind applying DP for IoV: how to synergistically combine DP with special characteristics of IoV, whose query sequences are usually formed as unbalanced batches in practice. The unique characteristic of IoV is the unbalanced sequence of requests (denoted as batches). These batches are formed due to a practical issue: vehicles require frequent queries of relevant data for in-time decision-makings~\cite{cidr-query}. Hence, how to strike a balance between privacy protection and data utility is a major challenge for applying DP for IoV \cite{UCC-TR/DP}. This is because privacy over-protection can cause low data utility. 

Prior works aim to balance the tradeoffs between data utility and privacy protection, but fail to address the issues of batches within IoV. \cite{Order} introduces Constraint Inference, which applies a constraint on the order of a sorted data set to decrease unnecessary noise; \cite{Micro-aggregation} introduces Micro-Aggregation, which uses a fixed-size cluster to cluster data set before adding noise, to reduce the information loss; and \cite{K-aggregation} introduces K-aggregation, which aggregates small-value data to form new data when the summation of small-value data exceeds a certain threshold. The ignorance prohibits DP to be applied effectively in the context of IoV, which demands new insights for bridging DP with IoV.

%However, this method fails to query on unit-length data, since all small-value data are aggregated as one new data and the process is irreversible.
%However, these two methods both lack generality on unevenly distributed dataset. 

We propose HUT, a new algorithm to enable \underline{H}igh \underline{UT}ility for DP-enabled protection in the context of IoV. Our key insight is to take advantage of the fact that, the interactions between centralized servers and edge vehicles are usually frequent and fine-grained, which are combined into a series of unbalanced batches. To exploit the characteristic, our key idea is to aggregate local batches and apply Order Constraints, so that information loss from DP protection can be mitigated.

%We propose HUT, a new DP-based protection algorithm to enhance the utility of the protected data. By utilizing the insights from aforementioned methods, we propose to use an adaptive algorithm, K-means algorithm  on small-value data in replace of fixed-size clustering on whole data set in Micro-Aggregation, together with Order Constraint mechanism to achieve high data utility.

We quantitatively evaluate the effectiveness of HUT against the state-of-the-art DP protection mechanism. Our results show that HUT can provide much lower information loss by 95.69\% and simultaneously enable strong mathematically-guaranteed protection of sensitive data. Meanwhile, HUT does not influence the strong guarantee of DP design, which ensures effective protection of sensitive data from users.

%In the following sections, we first present a thorough illustration of our algorithm in Section \ref{sec:overview}. Next, we introduce the experiment settings in Section \ref{sec:experiments} and present quantitative results of evaluation of our algorithm in Section \ref{sec:results}. Then we discuss future work and conclude our work in Section \ref{sec:conclusion}.

\section{Design Overview}
\label{sec:overview}
\begin{figure}[h]
  \centering
  \includegraphics[width=\linewidth]{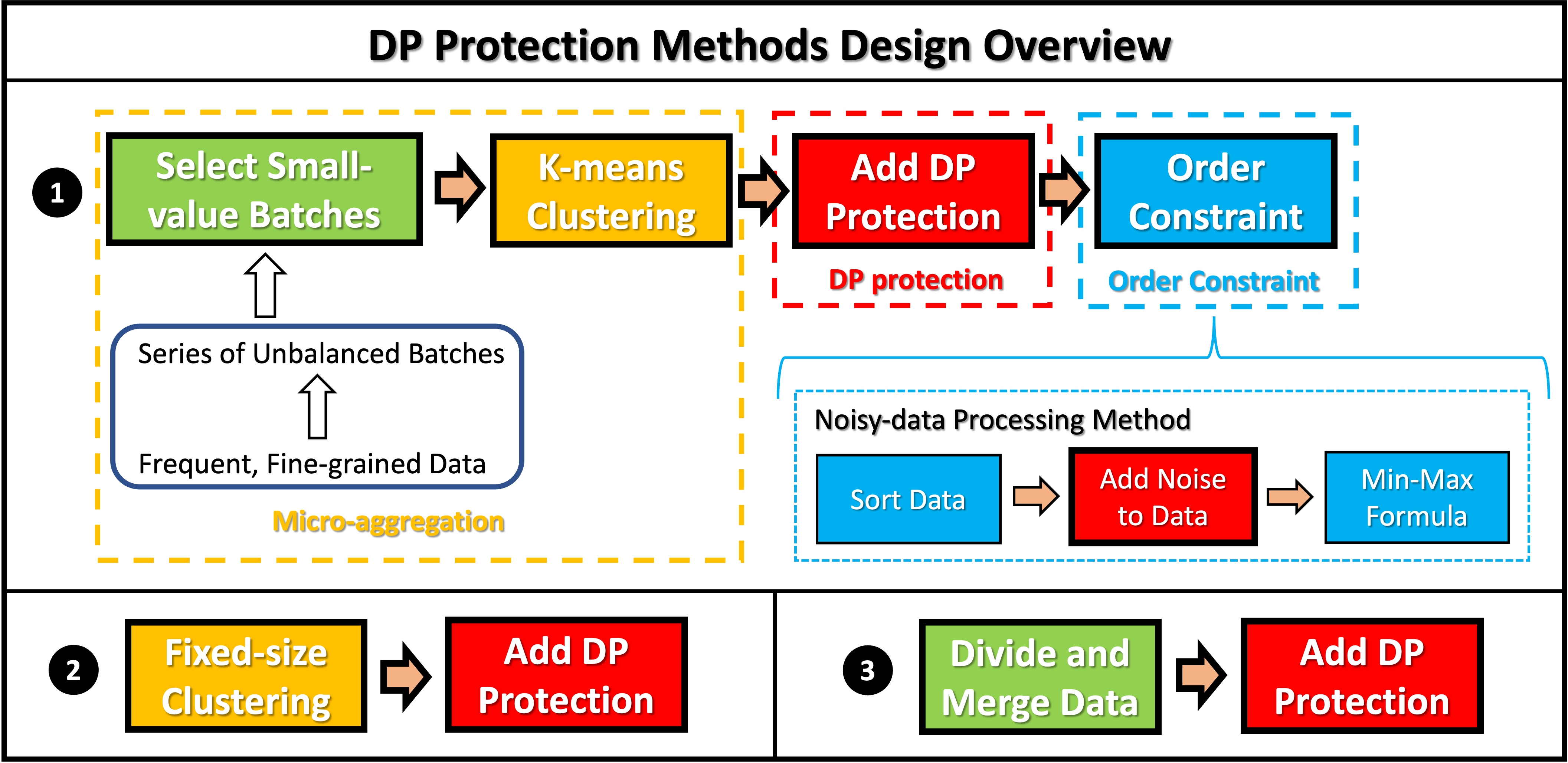}
  \caption{An illustration of three DP-protection methods. \ding{202} refers to HUT (our proposal), in which the details of the noisy-data processing method(OC)\cite{Order} is illustrated; Two state-of-the-art methods are illustrated (\ding{203} \cite{Micro-aggregation} and \ding{204} \cite{K-aggregation}).}
  \label{fig:comparison}
%  \vspace{2.0cm}
\end{figure}

We give an overview of our HUT design for ensuring DP protection with high data utility. HUT consists of three steps: 1) we leverage Micro-aggregation and K-means algorithm to group small-value batches together so that we can mitigate the negative effects of uniform DP protection; 2) we deploy Differential Privacy (DP) to preserve the privacy of the pre-processed data; 3) we exploit Order Constraint (OC) to re-sort DP-protected data so that we can achieve high data utility. Figure~\ref{fig:comparison} compares HUT with two other state-of-the-art DP protection methods, and we elaborate the differences in details as follows.

%Our method builds a DP-protected database with high data utility within three steps. First, we combine the advantages of Micro-aggregation and K-means algorithm to batch raw data with small values together for mitigating the negative effect of uniform DP protection (Step 1). Then, we utilize an advanced privacy-preserving technique, Differential Privacy (DP) to preserve the pre-processed data (Step 2). At last, we apply the insights from Order Constraint (OC) to re-sort DP-protected data and achieves high data utility (Step 3). Figure \ref{fig:comparison} illustrates the comparisons of our proposed methods and 3 other state-of-the art DP protection methods (details are discussed in the following sections).

%  \begin{figure*}[htbp]
%  \centering
%  \includegraphics[width=\linewidth]{img/Methods_comparison.png} 
%  \caption{A demo of the two different data set settings. The first situation is querying speed values from the Image-Speed Pair, while the second special situation is querying the counting numbers from Speed Value Counts.}
%  \label{fig:workload}
%  \end{figure*}
 
\noindent
1) \textbf{Micro-Aggregation:}
\label{subsec:step2}
HUT first applies Micro-Aggregation to cluster small-value raw data, so that we can mitigate the negative impacts from DP in terms of data utility. This is because DP adds uniform random noises to the dataset, and small-value data will have a higher Signal-to-Noise Ratio (SNR), compared with large-value data. Therefore, the data utility is impacted dependently according to its exact values. Therefore, a mechanism to mitigate such impacts is essential. We achieve so as follows. Before directly using DP techniques on raw data for protection, HUT sets a threshold to divide small-value data and the rest into different batches (as To-be-clustered batches and Not-clustered batches). Next, we apply K-Mean clustering on data within the threshold (the To-be-clustered batches), to automatically group such small-value data further into smaller intra-batches (i.e. Micro-Aggregation). 

Compared with \cite{Micro-aggregation}, HUT effectively reduces the sensitivity and variations of small-value data as they are upper bounded by the centroid of the respective cluster. In \cite{Micro-aggregation} (shown as \ding{203} in Figure \ref{fig:comparison}), a fixed-size clustering method is applied and this results in a consistent distance function on a total order relation but incurs large bias on the dataset which has highly uneven data distribution. Such bias on unevenly distributed data could harm the integrity of the dataset, leading to additional information loss. Thus, on the contrary, we leverage the K-means clustering algorithm to perform the clustering to mitigate the bias. \\ %\cite{K-aggregation} (shown as \ding{205} in Figure \ref{fig:comparison}) introduced the idea of using threshold to cluster small-value data from other data, but such data are then merged together into a single entry, which altered the structure of database and omit too much data set information.\\

\noindent
2) \textbf{DP Protection:}
\label{subsec:step3}
we then add Laplace noise on the data to achieve $\epsilon$-Differential Privacy. DP aims to minimize the changes in query answers when a record is changed in the data set by adding a randomized function (subject to $\epsilon$-differential privacy on the data). Hereby, we elaborate $\epsilon$-Differential Privacy via a mathematical description of a randomized function $N_f$ (shown below).

\noindent
\begin{theorem}
    
    (Randomized Function subject to $\epsilon$-DP) We regard a randomized function $N_f$ as satisfying $\epsilon$-DP, assuming that $\forall$ data set $X_1$ and $X_2$ with at most one element difference, and $\forall S \subseteq Range(K_f)$, the $N_f$ function obey the below in-equation:\\
    
    \centerline{$Pr[N_f(X_1) \in S] \leq exp(\epsilon) \times Pr[N_f(X_2) \in S]$}
\end{theorem}

The above-randomized function is commonly realized by adding Laplace Distributed Noise $\epsilon$, which is inversely proportional to the noise magnitude (If $\epsilon$ is smaller, the magnitude of the noise is larger). In HUT, we try out various $\epsilon$ settings to testify the proposed method feasibility in different protection scales.\\

\noindent
3) \textbf{Order Constraint:}
\label{subsec:step4}
% HUT finally applies Min-Max on the noisy data to constraint the order, which is introduced in the first step, to minimize unnecessary noise. In general cases, after adding noise on the previous sorted dataset, the noisy dataset is often disordered. \cite{Order} introduces OC to re-process the noisy data to ensure the same order as before (shown as \ding{203} in Figure \ref{fig:comparison}). Specifically, the Min-Max formulas (constraint inference) \cite{min-max} is used to  remove the unnecessary noise, applying least squares regression to change the data values in the noisy dataset to make sure it still follows the previous sorted order. The removed unnecessary noise is proved to have no influence on the effect of privacy protection but can have significant improvement on reducing information loss.\\
HUT finally utilizes a noisy-data processing method, Order Constraint(OC), to minimize unnecessary noise. A general feature of noisy data is that after adding noise(especially large-scale noise) on a previously sorted dataset, the noisy dataset often turns out to be disordered. \cite{Order} introduces OC to re-process the noisy data to ensure the same order as before. %(shown as \ding{203} in Figure \ref{fig:comparison}).
Specifically, the Min-Max formulas (constraint inference) \cite{min-max} is used to remove the unnecessary noise, applying least squares regression to change the data values in the noisy dataset to make sure it still follows the previous sorted order. The removed noise is proved to be unnecessary and does not influence the effect of dataset privacy protection but can have a significant improvement on reducing unnecessary information loss. Thus, we apply this method under the context of DP protection as a step to further reducing unnecessarily-added DP noise while having no negative impacts on privacy protection itself.\\

\noindent
\textbf{Novelty of HUT:} HUT has three novelties, compared with the state-of-the-art DP protection mechanisms. First, HUT applies two-stage Micro-aggregation to mitigate the negative impacts on fine-grained batches, which is considered as a unique characteristic of IoV; Second, HUT leverages K-Means Clustering, rather than fixed-size clustering, to improve the utility of protected data; and third, HUT selectively enables lightweight constraints to be synergistic with other designs, for both DP protection and data utility .\\

\noindent
\textbf{Are there any negative impacts on DP protection:}
In short, there are no negative impacts on DP protection when applying HUT: since DP protection and K-means clustering are not in conflict with the mathematical guarantee of DP, we focus on the Micro-Aggregation algorithm. First, Micro-Aggregation in HUT does not impact the DP protection since K-Means Clustering constrains only small-valued data; and second, Micro-Aggregation in HUT does not change the data, and therefore no negative impacts are incurred on DP protection.

 \begin{figure*}[htbp]
 \centering
 \includegraphics[width=.85\linewidth]{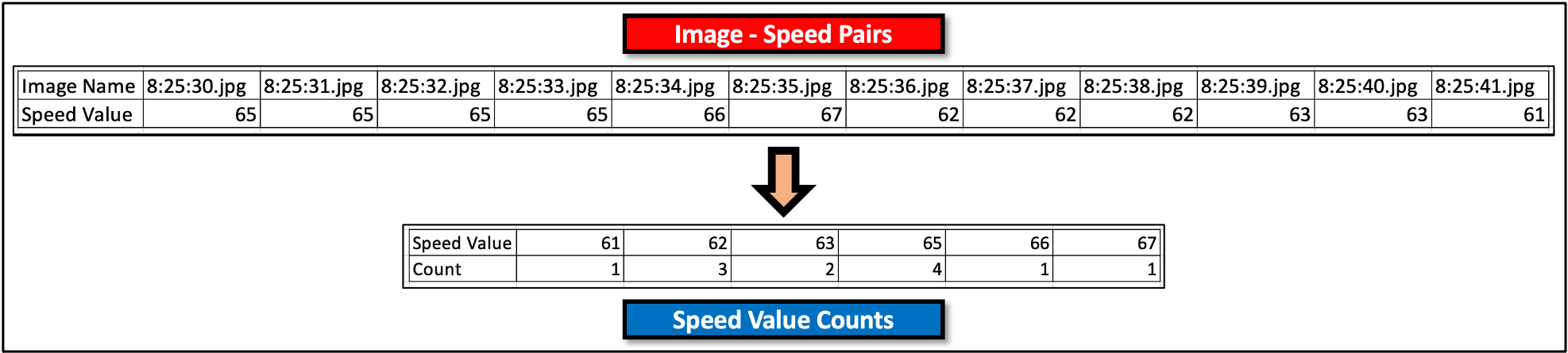} 
 \caption{A demo of the two different data set settings. The first situation is querying speed values from the Image-Speed Pair which simulates Simple Query, while the second special situation is querying the counting numbers from Speed Value Counts, which simulates Counting Query.}
 \label{fig:workload}
 \end{figure*}

\section{Experimental Setup}
\label{sec:experiments}
\vspace{-2pt}
$\bullet$ \textbf{Selected Dataset:} 
% \\ 1. BROOK overview 2. hand crafted dataset from BROOK (face-img, speed) pair
% We conduct our experiments on the BROOK dataset, which is multi-modal database that using facial video record frames to characterize drivers driving status. In our experiments, we specifically hand crafted a dataset that contains image-speed pair text data (part of this dataset is shown in []). Each image-speed pair consists an image name (served as label) and a speed value corresponding to the image name.
We use BROOK dataset \cite{BROOK}. BROOK dataset is a multi-modal dataset with facial videos and 11 dimensions of biosignal/contextual data from 34 drivers' driving experiences. In our experiments, we specifically handcraft a dataset that contains drivers' facial images and the corresponding vehicle speeds, as a proof-of-concept. Figure~\ref{fig:workload} demonstrates an example of our handcrafted dataset.\\ %where the jpg files are drivers' facial images (images are named corresponding to the capture time, e.g: "8:25:30.jpg" represents a picture captured at 8:25am 30 seconds.) and the numbers below image files are the corresponding vehicle speed for each image.

\noindent
\vspace{-2pt}
$\bullet$ \textbf{Configurations of Proposed Methods:}
There are three key parameters for our method, the scale of DP $\epsilon$, the threshold $p$ and the number of clusters $k$ for Micro-Aggregation. We testify our design in various parameter settings, including $\epsilon \in \{0.008, 0.01, 0.02, 0.05\}$, $k  \in \{5, 8, 10, 15\}$ and $p \in \{30$\%$, 35$\%$, 40$\%$\}$, in which $k  \in \{5, 8, 10\}$ are used for simple queries and $k  \in \{5, 10, 15\}$ are used for counting queries to satisfy different data set configurations. We conduct 20 experiments for each parameter setting and take the mean as the final value of information loss.\\ %We compared the information loss of our method with Raw DP protection as validation.\\

\noindent
\vspace{-2pt}
$\bullet$ \textbf{Query Settings:} 
We examine the data utility of the DP-protected dataset with two representative types of queries. The first query type is a simple query with facial images as input, and a query for the speed data in our handcrafted dataset. These queries simulate the unit-length querying process for intelligent in-vehicle systems to continuously retrieve data for decision-making in fine granularity. 
%In this scenario, the sensitivity of original data is high and accumulations of query errors can significantly poison the in-vehicle systems for making irrational decisions. Therefore, it is vital to testify whether our algorithm can decrease information loss more effectively. 
The second type of query is the counting query, which simulates the statistical analysis function of in-vehicle systems. The counting query is similar to query data from a contingency table (the bottom image in Figure \ref{fig:workload}). Different from the first general situation, this one already has a fixed sensitivity of 1, making it more challenging for HUT because sensitivity reduction is limited.\\

\noindent
\vspace{-2pt}
$\bullet$ \textbf{Evaluation Metric:} We use the Mean Square Error (MSE) as the evaluation metric of data utility and information loss. We denote the query response on the DP-protected data as $d'$, and the query response on the raw data as $d$. MSE is calculated by equation $\mathrm{MSE}=\frac{1}{n} \sum_{i=1}^{n}\left(d'_{i}-{d}_{i}\right)^{2}$.

% \begin{figure*}[H]
% \begin{minipage}{0.5\linewidth}
%   \centering
%   \includegraphics[width=\linewidth]{img/simple_epsilon2.png}
%   \caption{Average information loss of different epsilon values for simple queries (when k = 10, threshold = 35$\%$). }
%   \label{fig:simple_epsilon}
% %  \vspace{2.0cm}
% \end{minipage}
% \hfill
% \begin{minipage}{0.5\linewidth}
%   \centering
%   \includegraphics[width=\linewidth]{img/count_epsilon2.png}
%   \caption{Average information loss of different epsilon values for counting queries (when k = 10, threshold = 35$\%$). }
%   \label{counting_epsilon}
% %  \vspace{2.0cm}
% \end{minipage}
% \end{figure*}

\section{Experimental Results}
\label{sec:results}

We compare HUT with the state-of-the-art DP algorithms to examine its effectiveness of data utility. Since the average information loss of method \ding{203} and \ding{204} in Figure~\ref{fig:comparison} are highly similar under all experiment conditions, we select the overall best performer \ding{204} to compare its information loss with the information loss of HUT.

\begin{figure}[h]
  \centering
  \includegraphics[width=.7\linewidth]{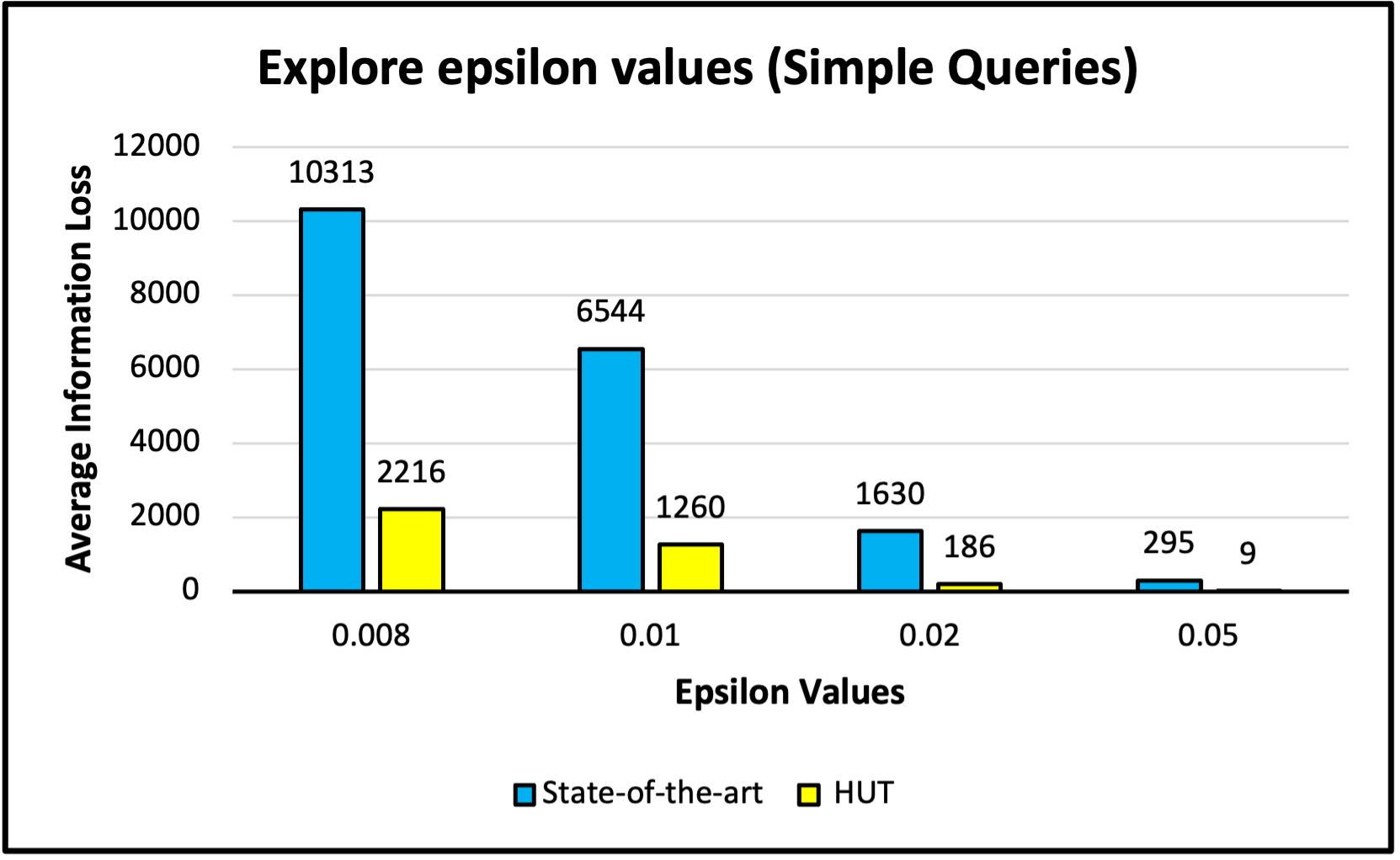}
  \caption{Average information loss of different epsilon values for simple queries (when k = 10, threshold = 35$\%$). }
  \label{fig:simple_epsilon}
%  \vspace{2.0cm}
\end{figure}

\begin{figure}[htbp]
    \centering

  \includegraphics[width=.7\linewidth]{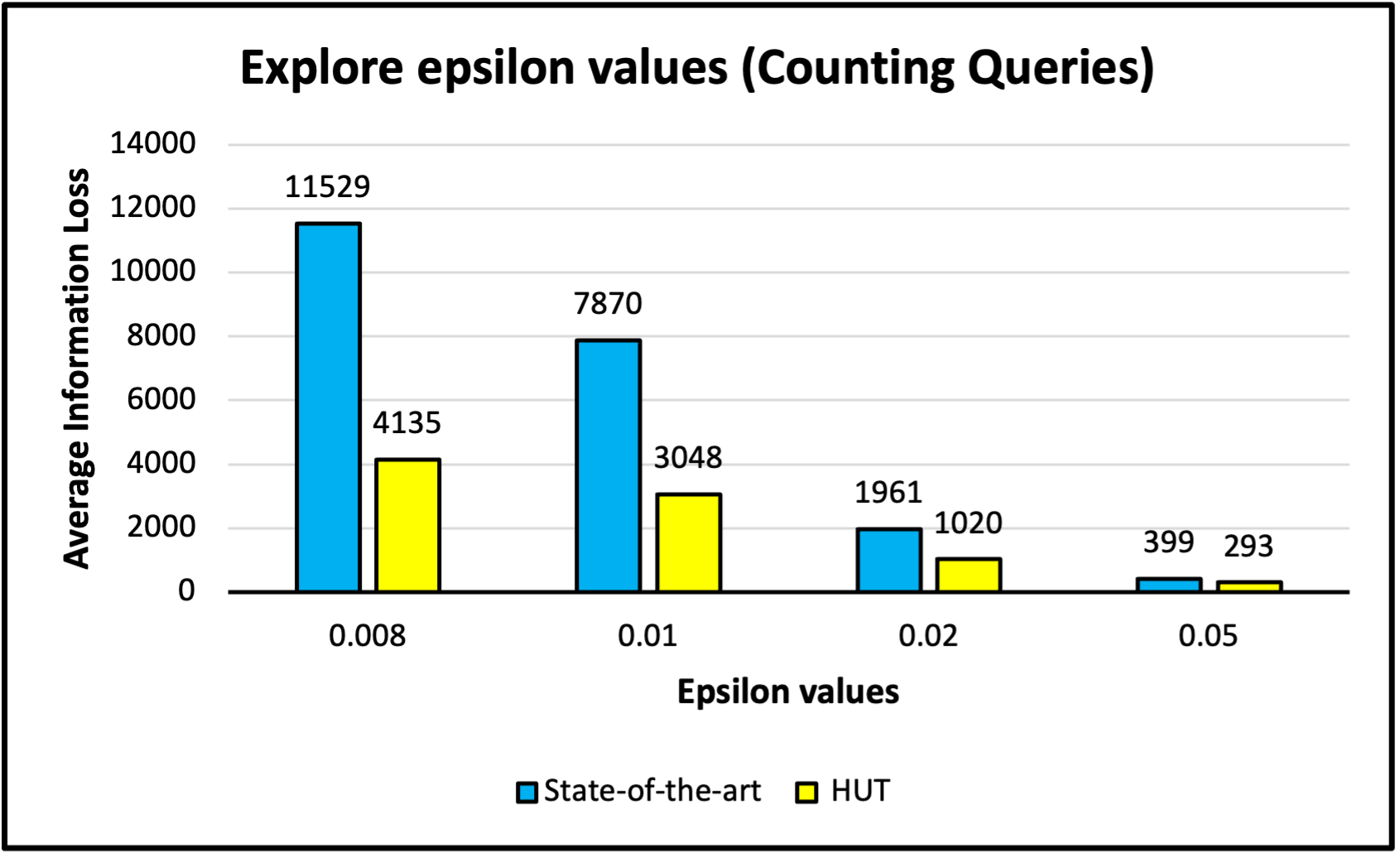}
  \caption{Average information loss of different epsilon values for counting queries (when k = 10, threshold = 35$\%$). }
  \label{counting_epsilon}
%  \vspace{2.0cm}
\end{figure}

 The results of the simple query and counting query in different $\epsilon$ values with fixed k clusters and threshold portions are presented in Figure~\ref{fig:simple_epsilon} and Figure~\ref{counting_epsilon}. Both figures prove the effectiveness of Micro-aggregation in increasing SNR for small-value data and mitigating the negative effects of DP protections, in a common setting (simple query) and challenging setting (counting query). For simple query, our method achieves a maximum of 98\%  (when $\epsilon$=0.008) and a minimum of 79\% (when $\epsilon$=0.05)  reduction information loss reduction compared to the state-of-the-art, with k=10 and 30\% threshold. For the counting query, the max information loss reduction percentage is 65\% (when $\epsilon$=0.008) and the minimum percentage is 27\% (when $\epsilon$=0.05). We note that when the noise magnitude grows (i.e. $\epsilon$ values decreases), the performance of our algorithm degrades in a simple query, but improves in counting query.  % explanation for the opposite trend??
 The opposite trend in two query settings proves two things: 1) Applying DP on raw data could significantly harm normal data utility, as simple query, a primary indexing operation could cause relatively severe information loss; 2) Aggregation, as one kind of counting operations, could effectively avoid over-protection in extreme conditions, but is less effective when data is not over-protected.


\begin{figure}[h]
  \centering
  \includegraphics[width=.7\linewidth]{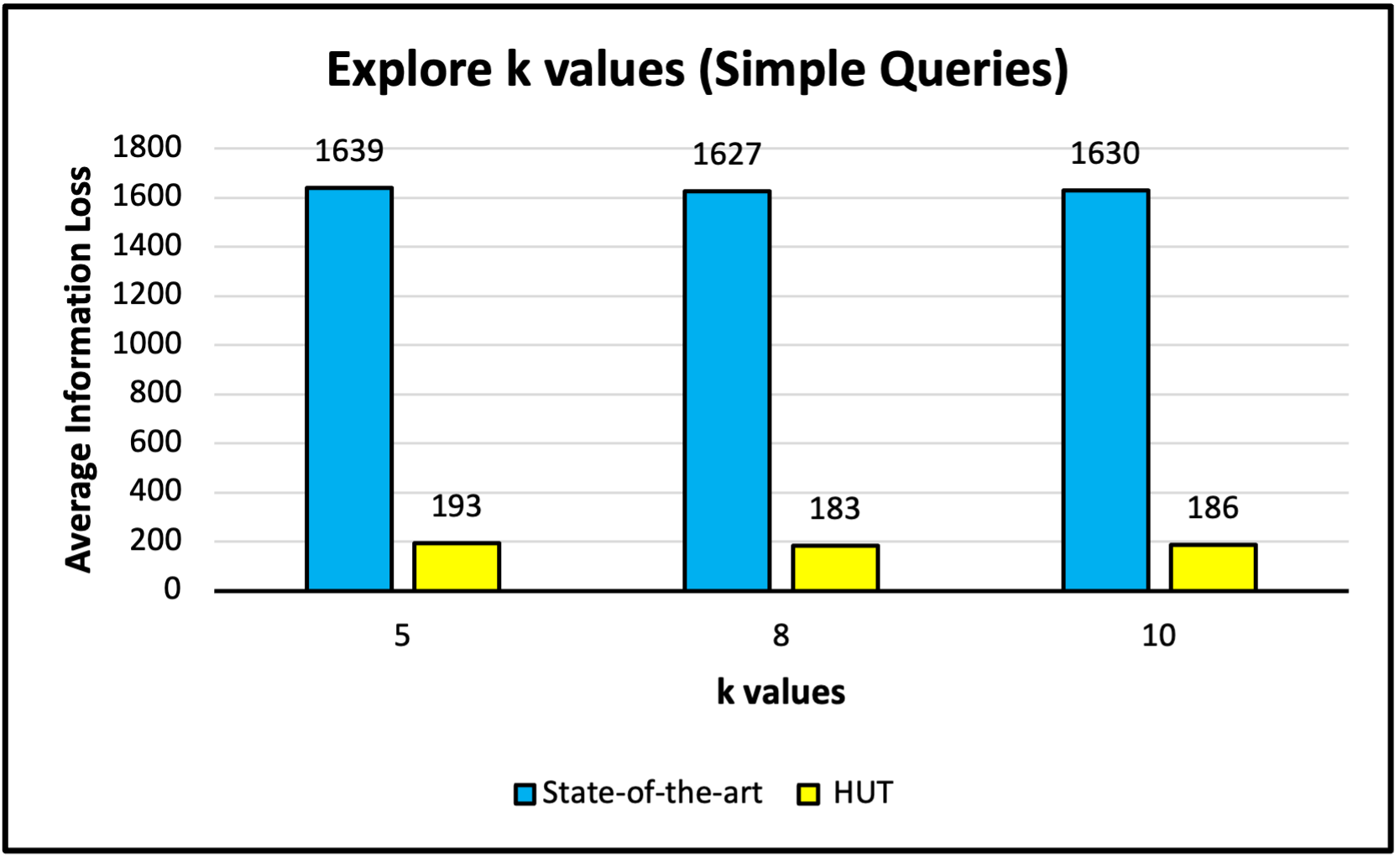}
  \caption{Average information loss of different k values for simple queries (when epsilon = 0.02, threshold = 35$\%$). }
  \label{fig:simple_k}
%  \vspace{2.0cm}
\end{figure}

\begin{figure}[h]
    \centering
  \includegraphics[width=.7\linewidth]{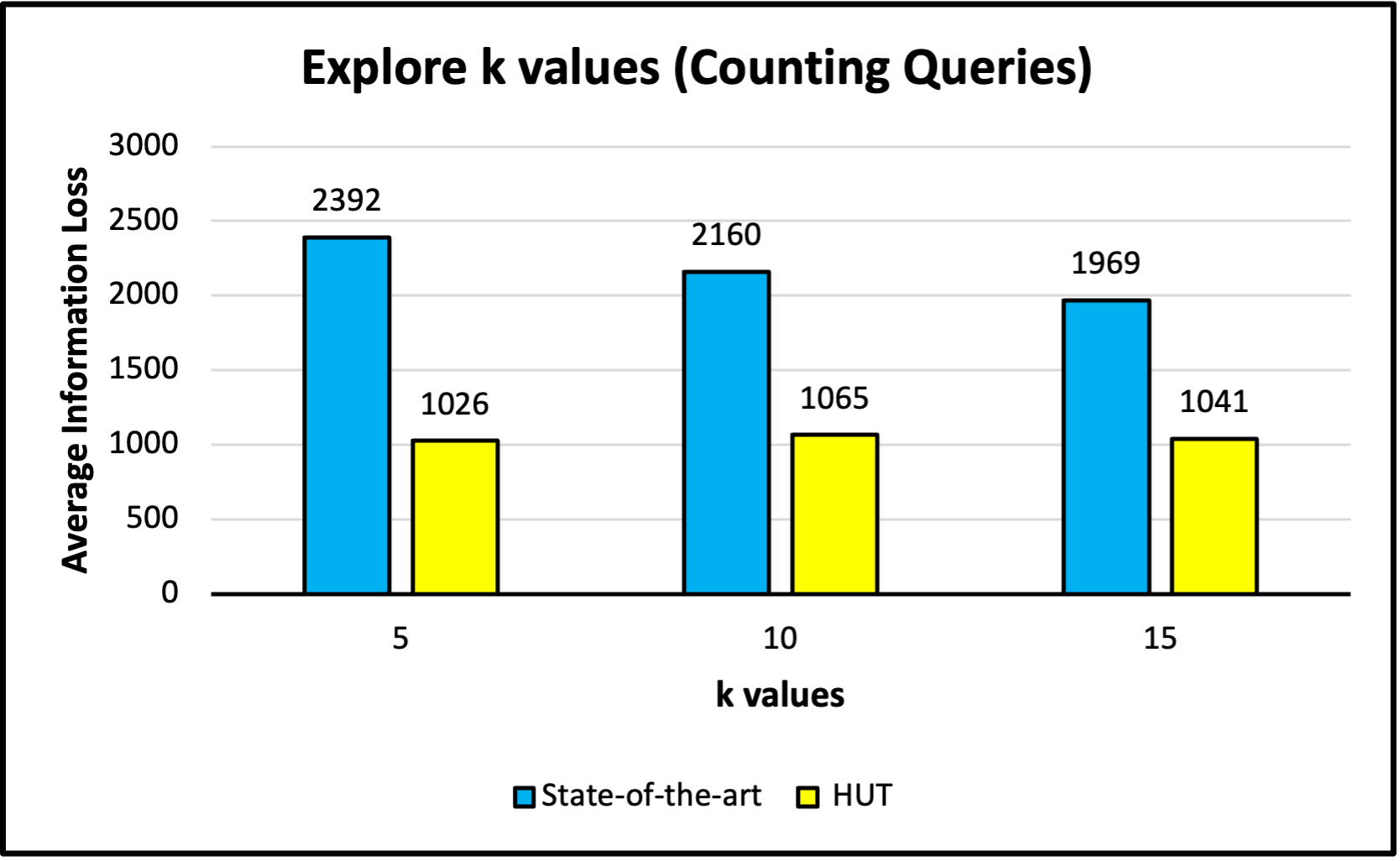}
  \caption{Average information loss of different k values for counting queries (when epsilon = 0.02, threshold = 35$\%$). }
  \label{fig:counting_k}
%  \vspace{2.0cm}
\end{figure}

% choices of k and threshold
Figure~\ref{fig:simple_k} and Figure~\ref{fig:counting_k} presents the information loss of simple query and counting query in different K settings, when $\epsilon$ and threshold are fixed at 0.02,35\%. We obtain the key observations that an effective choice of the number of clusters for K-mean clustering falls in the range between 5 and 15, which have the optimal information loss reduction. In addition, our empirical study shows that any reasonable threshold values would incur negligible perturbations on the information loss, as different threshold values cause a maximum 1\% variance for simple query and 6\% for counting query.

Table~\ref{tab:best} reports the best percentage of average information loss compared to the state-of-the-art mechanism in both simple query and counting query for each $\epsilon$ value when the number of clusters k and threshold for aggregation varies for each selected $\epsilon$. We believe exploiting the mitigation of DP's negative effects in different $\epsilon$ settings is essential, as a small scale of $\epsilon$ changes could cause relatively large fluctuations in the information loss.

\begin{table}[!h]
\caption{Best information loss reduction on the state-of-the-art mechanism in different $\epsilon$ settings}
\centering
\begin{tabular}{ccccc}
\hline
                                                                           & Epsilon & K & Threshold & \% of reduction \\ \hline
\multirow{4}{*}{\begin{tabular}[c]{@{}c@{}}Simple\\  Query\end{tabular}}   & 0.008   & 10             & 30\%      & 78.51\%         \\ \cline{2-5} 
                                                                           & 0.01    & 8              & 30\%      & 80.74\%         \\ \cline{2-5} 
                                                                           & 0.02    & 8              & 30\%      & 88.75\%         \\ \cline{2-5} 
                                                                           & 0.05    & 5              & 30\%      & \textbf{95.69\%}         \\ \hline
\multirow{4}{*}{\begin{tabular}[c]{@{}c@{}}Counting \\ Query\end{tabular}} & 0.008   & 10             & 35\%      & 64.13\%         \\ \cline{2-5} 
                                                                           & 0.01    & 5              & 30\%      & \textbf{71.71\% }        \\ \cline{2-5} 
                                                                           & 0.02    & 5              & 35\%      & 45.22\%         \\ \cline{2-5} 
                                                                           & 0.05    & 10             & 40\%      & 50.46\%         \\ \hline
\end{tabular}
\label{tab:best}
\end{table}

\section{Conclusions and Future Works}
\label{sec:conclusion}

We present HUT, a new algorithm to enable \underline{H}igh \underline{UT}ility for DP-enabled protection in the context of IoV. Our key insight is to take advantage of the fact that, the interactions between centralized servers and edge vehicles are usually frequent and fine-grained, which are combined into a series of unbalanced batches. We evaluate the effectiveness of HUT against the state-of-the-art DP protection mechanism, and the results show that HUT can provide much lower information loss by 95.69\% and simultaneously enable strong mathematically-guaranteed protection of sensitive data.

Our future works aim to explore the scalability and robustness of HUT. For scalability, we expect to examine whether HUT can be applied for data with broad range or unevenly distributed datasets; as for robustness, HUT still suffers from the instability of the K-mean clustering method, which can produce query results with relatively large variance. Therefore, we can expect a trade-off between different clustering algorithms, in terms of the robustness of HUT.

\bibliographystyle{IEEEbib}
\bibliography{refs}
\end{document}